\documentclass[pre,twocolumn,showpacs]{revtex4}

\usepackage{graphicx}

\begin{document}

\newcommand{\rbc}{Rayleigh-B\'enard convection}
\newcommand{\dotprod}{\,{\scriptscriptstyle \stackrel{\bullet}{{}}}\,}
\newcommand{\gtapprox}{\,{\scriptscriptstyle \stackrel{>}{\sim}}\,}
\newcommand{\ltapprox}{\,{\scriptscriptstyle \stackrel{<}{\sim}}\,}
\renewcommand{\vec}{\mathbf}
\newcommand{\figwidth}{3in}
\newcommand{\widefigwidth}{4.5in}
\newcommand{\pec}{\mathcal{P}}
\newcommand{\leff}{\mathcal{L}^\ast}
\newcommand{\lew}{\mathcal{L}}

\title{Enhanced tracer transport by the spiral defect chaos state of a
  convecting fluid}

\author{K.-H.~Chiam}
\email{ChiamKH@MailAPS.ORG}
\homepage{http://www.cmp.caltech.edu/~stchaos}

\author{M.~C.~Cross}
\affiliation{Nonlinear and Statistical Physics, Mail Code 114-36,
  California Institute of Technology, Pasadena, CA 91125-3600}

\author{H.~S.~Greenside}
\affiliation{Department of Physics, P.O.~Box 90305, Duke University,
  Durham, NC 27708-0305}

\author{P.~F.~Fischer}
\affiliation{Mathematics and Computer Science Division, Argonne
National Laboratory, Argonne, IL 60439}

\date{\today}

\begin{abstract}
  To understand how spatiotemporal chaos may modify material
  transport, we use direct numerical simulations of the
  three-dimensional Boussinesq equations and of an advection-diffusion
  equation to study the transport of a passive tracer by the spiral
  defect chaos state of a convecting fluid.  The simulations show that
  the transport is diffusive and is enhanced by the spatiotemporal
  chaos.  The enhancement in tracer diffusivity follows two regimes.
  For large P\'eclet numbers (that is, small molecular diffusivities
  of the tracer), we find that the enhancement is proportional to the
  P\'eclet number.  For small P\'eclet numbers, the enhancement is
  proportional to the square root of the P\'eclet number.  We explain
  the presence of these two regimes in terms of how the local
  transport depends on the local wave numbers of the convection rolls.
  For large P\'eclet numbers, we further find that defects cause the
  tracer diffusivity to be enhanced locally in the direction
  orthogonal to the local wave vector but suppressed in the direction
  of the local wave vector.
\end{abstract}


\maketitle

\section{Introduction}
\label{se:intro}
This paper addresses the transport of passive neutrally buoyant
tracers in \rbc\ exhibiting spiral defect chaos --- an example of
spatiotemporal chaos that is characterized by disorder in both space
and time~\cite{Cross94,Gollub00,Hohenberg89}.  An important
characteristic of such spatially disordered flows is that fluctuations
in space play a significant role in their dynamics, resulting in
advection of the passive tracers that is dependent in a complex
fashion on space and time.  The transport of passive tracers in such
disordered flows is then governed by this advection in addition to
molecular diffusion.  The goal of this paper is to understand the net
average transport of passive tracers as a function of the two
competing mechanisms of advection by spatiotemporal chaos and
molecular diffusion.  Understanding material transport by
spatiotemporal chaos is a problem that is of considerable importance
in many branches of science and engineering.  For example, an improved
understanding may allow one to gain insight into heat and mass
transport in atmospheric and oceanic flows and also in chemical
engineering processes such as combustion.

Previous studies of the properties of passive transport in convective
flows have focused only on the steady and weakly oscillatory regimes.
For example, in two-dimensional time-independent laminar \rbc\ flow,
experiments have shown that the transport is effectively diffusive in
the long time limit, with an effective diffusivity that is greater
than the molecular diffusivity by a factor that scales as the square
root of the P\'eclet number (defined in Eq.~(\ref{eq:peclet}) to be
the ratio of the strength of advection to diffusion) \cite{Solomon88}.
This enhancement, in the large P\'eclet number limit, has also been
calculated theoretically by using the matched asymptotic expansion
method \cite{Rosenbluth87,Shraiman87}.  In addition, higher-order
corrections to the diffusion process, for arbitrary P\'eclet numbers,
have been calculated numerically using the homogenization method
\cite{McCarty88,McLaughlin85}.  For nearly two-dimensional
time-periodic convection, experiments near the onset of the
oscillatory instability \cite{Busse78} have shown that the transport
is again effectively diffusive but with an effective diffusivity that
depends linearly on the local amplitude of the roll oscillations
\cite{Solomon88b}.  This result has also been confirmed in theoretical
work, which also identified the invariant structures of the flow that
acted as templates for the motion of the tracers
\cite{Camassa91,Camassa91b}.  Passive tracer transport has also been
studied in other types of laminar flow, including capillary waves
generated by the Faraday instability \cite{Ramshankar90,Ramshankar91}
and Taylor-Couette flow in a rotating annulus
\cite{Solomon93,Solomon94}.

In this paper, the above transport studies are extended to flows that
exhibit spatiotemporal chaos.  It will be shown that the transport is
globally diffusive and is enhanced by the spatiotemporal chaos.
However, unlike the case of laminar flows, the enhancement is found to
follow two regimes.  For large P\'eclet numbers (that is, small
molecular diffusivities of the tracer), the enhancement is
proportional to the P\'eclet number, whereas for small P\'eclet
numbers, the enhancement is proportional to the square root of the
P\'eclet number.  These two regimes are then explained by analyzing
how the local transport depends on the local wave numbers of the
convection rolls.

The remainder of this paper is organized as follows. In
Sect.~\ref{se:def}, the equations governing \rbc\ and the transport of
passive tracers are defined.  In addition, direct numerical
simulations of these equations are discussed.  In Sect.~\ref{se:res},
results from these simulations are presented.  In Sect.~\ref{se:con},
conclusions are presented.

\section{Equations and Algorithms}
\label{se:def}
\subsection{Rayleigh-B\'enard Convection}

\begin{figure}
  \begin{center}
    \includegraphics[width=\figwidth]{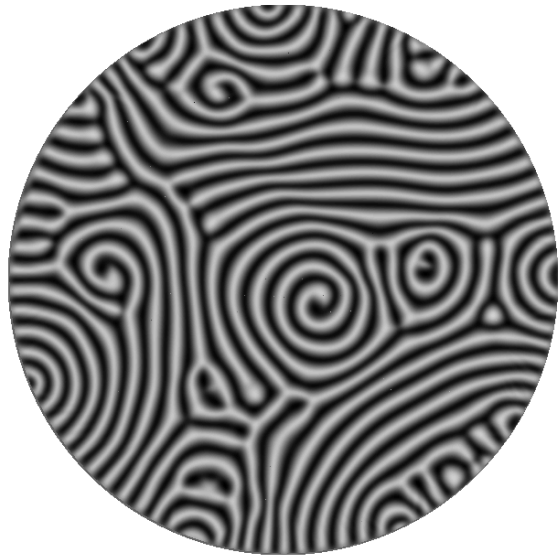}
  \end{center}
  \caption{Example of spiral defect chaos observed in a numerical
    simulation described in Sect.~\ref{se:dns} with insulating and
    no-slip boundaries, and with a spatial resolution of $\Delta x =
    1/8$ and a temporal resolution of $\Delta t = 10^{-3}$.  The
    mid-plane temperature field is plotted at time $t=500$ for
    parameters $\epsilon=1.0$, $\sigma=1$, and in a cylindrical
    geometry of aspect ratio $\Gamma=30$.  Dark regions correspond to
    cold sinking fluid, light regions to hot rising fluid.  The spiral
    defect chaos planform is characterized by a disordered collection
    of spirals rotating in both directions and coexisting with
    dynamical defects such as grain boundaries and dislocations.}
  \label{fig:sdc}
\end{figure}

In a typical \rbc\ experiment, an incompressible fluid layer is
confined between two horizontal plates, and is thermally driven far
from equilibrium by maintaining the bottom plate at a temperature that
is higher than that of the top plate.  As the temperature difference
is increased, the fluid undergoes an instability to a state in which
there is motion driven by the buoyancy forces. When the temperature
difference between the plates is above but near this convective
threshold, a pattern comprising patches of locally parallel convection
rolls forms with roll diameters that are close to the depth of the
cell.  When the temperature difference is increased, the fluid
undergoes other instabilities that may result in the pattern
developing an oscillatory or chaotic time-dependence.  Finally when
the temperature difference is increased further and if the aspect
ratio~$\Gamma$ is larger than about~$20$ in boxes and~$30$ in
cylinders, spiral defect chaos appears.  This state is a disordered
collection of spirals that rotate in both directions and coexist with
dynamical defects such as grain boundaries and dislocations.
Fig.~\ref{fig:sdc} shows a numerically simulated instance of the
spiral defect chaos state in a cylindrical geometry.  More generally,
spiral defect chaos is an example of a kind of widely observed
phenomenon called spatiotemporal chaos that exhibits disorder in space
and chaos in time.

The evolution of the convecting fluid is governed to good
approximation by the three-dimensional Boussinesq equations
\cite{Cross93}.  They are the combination of the incompressible
Navier-Stokes and heat equations, with the further assumption that
density variations are proportional to temperature variations and that
this density variation appears only in the buoyancy force.  Written in
a dimensionless form, they are:
\begin{eqnarray}
\label{eq:NS}
\sigma^{-1} \left( \partial _{t}+\vec{u}\dotprod \vec{\nabla}
\right) \vec{u}(x,y,z,t) &=&-\vec{\nabla }p+ \vec{\nabla }^{2}\vec{u}+
RT\widehat{z}, \\
\label{eq:T}
\left( \partial _{t}+\vec{u}\dotprod \vec{\nabla} \right) T(x,y,z,t)
&=&\vec{\nabla }^{2}T, \\
\label{eq:incom} \vec{\nabla} \dotprod \vec{u} &=&0.
\end{eqnarray}
The field $\vec{u}(x,y,z,t)$ is the velocity field at point $(x,y,z)$
at time $t$, while $p$ and $T$ are the pressure and temperature
fields, respectively.  The variables $x$ and $y$ denote the horizontal
coordinates, while the variable $z$ denotes the vertical coordinate,
with the unit vector $\widehat{z}$ pointing in the direction opposite
to the gravitational acceleration. The spatial units are measured in
units of the cell depth $d$, and time is measured in units of the
vertical thermal diffusion time $d^2/\kappa$, where $\kappa$ is the
thermal diffusivity of the fluid.  The parameter $R$ is the Rayleigh
number, a dimensionless measure of the temperature difference $\Delta
T$ across the top and bottom plates,
\begin{equation}
  \label{eq:R}
  R = \frac{\alpha g d^3}{\nu \kappa} \Delta T,
\end{equation}
where $\alpha$ is the thermal expansion coefficient, $\kappa$ the
thermal diffusivity, and $\nu$ the viscous diffusivity (kinematic
viscosity) of the fluid. In this paper, the reduced Rayleigh number
will also be frequently used,
\begin{equation}\label{eq:reducedR}
  \epsilon = \frac{R-R_c}{R_c},
\end{equation}
where $R_c \approx 1708$ is the critical Rayleigh number at the onset
of convection in an infinite domain \cite{Cross93}.  The parameter
$\sigma$ is the Prandtl number, defined to be the ratio of the fluid's
thermal to viscous diffusivities,
\begin{equation}
  \label{eq:sigma}
  \sigma = \frac{\nu}{\kappa}.
\end{equation}

The material walls are no-slip so that the velocity field satisfies
\begin{equation}
    \label{eq:u_bcs}
  \vec{u} = \vec{0}, \ \ \ \mbox{on all material walls}.
\end{equation}
The temperature field is constant on the top and bottom plates:
\begin{equation}
    \label{eq:T_tb_bcs}
  T\left(x,y,z=\mp\frac{1}{2},t\right)=\pm \frac{1}{2}.
\end{equation}
The lateral walls are assumed to be perfectly insulating, so that
\begin{equation}
  \label{eq:T_lateral_bcs}
  \widehat{n}\dotprod \vec{\nabla} T = 0, \ \ \ \mbox{on lateral walls,}
\end{equation}
where $\widehat{n}$ is the unit vector perpendicular to the lateral
walls at a given point.  The pressure field $p$ has no associated
boundary condition because it does not satisfy a dynamical equation.

The influence of the lateral walls on the dynamics is determined by
the dimensionless aspect ratio $\Gamma$, defined to be the
half-width-to-depth ratio of the cell if it is rectangular and the
radius-to-depth ratio if it is cylindrical.

\subsection{Transport Equation}

The transport of passive neutrally buoyant tracers in a flow can be
described by the advection-diffusion equation.  Written in a
dimensionless form, it is
\begin{equation}
  \label{eq:tpt}
  (\partial_t + \vec{u}\dotprod\vec{\nabla}) \psi(x,y,z,t) = \lew
  \vec{\nabla}^2 \psi.
\end{equation}
The scalar field $\psi(x,y,z,t)$ is the passive tracer concentration
at point $(x,y,z)$ and time $t$.  The velocity field $\vec{u}$ is
obtained by solving the Boussinesq equations,
Eqs.~(\ref{eq:NS})--(\ref{eq:incom}).  The parameter $\lew$ is the
Lewis number, defined to be the molecular diffusivity $D$ of the
tracers made dimensionless by the thermal diffusivity $\kappa$ of the
fluid,
\begin{equation}
  \lew = \frac{D}{\kappa}.
\end{equation}
In this paper, small Lewis numbers in the range $10^{-3} \le \lew \le
10^{-1}$ will be considered.  In comparison, the Lewis numbers of
passive tracers used in previous convection
experiments~\cite{Solomon88} in water at approximately $300$~K,
namely, micrometer-sized latex spheres (vinyl toluene
$t$-butylstyrene) and methylene blue dye, are $\lew=1.2\times 10^{-5}$
and $\lew=3.9\times 10^{-3}$, respectively.

The tracers are assumed to be passive, that is, their motions in the
fluid do not modify the fluid's velocity field.  The fluid is also
assumed to have negligible Soret and Dufour effects.  The former
refers to the additional passive tracer concentration current driven
by gradients of the temperature field, whereas the latter refers to
the additional heat current driven by gradients of the passive tracer
concentration.  In addition, the lateral walls are assumed to be
impermeable to the tracers, so that
\begin{equation}
  \label{eq:psi_lateral_bcs}
  \widehat{n}\dotprod \vec{\nabla} \psi = 0 \ \ \ \mbox{on lateral walls,}
\end{equation}
where $\widehat{n}$ is the unit vector perpendicular to the lateral
walls at a given point.

Eq.~(\ref{eq:tpt}) is also commonly written in the literature in an
alternate but entirely equivalent form.  By dividing it throughout by
the product of a characteristic velocity scale $||\vec{u}||$ and a
characteristic length scale, Eq.~(\ref{eq:tpt}) becomes
\begin{equation}
  ( \partial_{\tilde{t}} + \vec{\tilde{u}}\dotprod \vec{\nabla}) \psi(x,y,z,t) =
  \frac{1}{\pec} \vec{\nabla}^2 \psi,
\end{equation}
with $\tilde{t}$ the rescaled time, $\vec{\tilde{u}}$ the rescaled
velocity field, and $\pec$ the P\'eclet number, defined to be the
dimensionless ratio of the relative importance of the advection of the
tracers to their molecular diffusion,
\begin{equation}
  \label{eq:peclet}
  \pec = \frac{||\vec{u}||}{\lew}.
\end{equation}
(Note that the numerator in Eq.~(\ref{eq:peclet}) contains a
characteristic length scale --- the depth of the cell --- which is
unity and is thus omitted.)

Finally, it should also be noted that instead of studying the passive
tracer concentration field $\psi$ in the space coordinates defined in
the laboratory frame (the Eulerian approach), one could study the
trajectories of each passive tracer individually (the Lagrangian
approach) by integrating, for each passive tracer,
\begin{equation}
  \label{eq:lagrangian}
  \frac{d\vec{x}(t)}{dt} = \vec{u}[\vec{x}(t),t]  + \vec{\eta}(t),
\end{equation}
where $\vec{x}(t)$ is the position of the tracer [initially at
$\vec{x}(0)$], $\vec{u}$ is the Eulerian velocity field at space
$\vec{x}(t)$ and time $t$, and $\vec{\eta}(t)$ is a Langevin noise
introduced to represent molecular diffusion.  However, this approach
is not pursued here because of the difficulties associated with
integrating Eq.~(\ref{eq:lagrangian}).  In fact, even if the velocity
field $\vec{u}$ can be explicitly determined and has a very simple
form, the tracer trajectories $\vec{x}$ can have very complicated
dynamics \cite{Aref83}.

\subsection{Direct Numerical Simulations}
\label{se:dns}

We use a parallel spectral element scheme to integrate the Boussinesq
equations, Eqs.~(\ref{eq:NS})--(\ref{eq:incom}), and the transport
equation, Eq.~(\ref{eq:tpt}).  The scheme is second-order-accurate in
time and is designed for rectangular, cylindrical, as well as more
complex geometries with arbitrary lateral boundary conditions.
Details of this scheme are available elsewhere \cite{Fischer97}.  For
applications of this scheme to related problems in \rbc, see
Refs.~\cite{Paul03,Chiam03,Paul01,Paul02,Paul04}.

For small Lewis numbers $\lew \ll 1$, one well-known
difficulty~\cite{Fletcher91,Tannehill97} associated with integrating
Eq.~(\ref{eq:tpt}) is that the spatial resolution $\Delta x$ has to be
very small.  This scale is set by the smallest scale in the tracer
concentration field, such as the thickness of the interface where the
tracer is initially zero on one side and unity on the other.  The
interface is then stretched by a strain rate $S \sim \partial
u/\partial x \sim ||\vec{u}||$, and its thickness is proportional to
$(\lew/S)^{1/2}$, that is,
\begin{equation}
  \label{eq:mesh-reynolds}
  \Delta x \sim \left( \frac{\lew}{||\vec{u}||} \right)^{1/2} = \pec^{-1/2}.
\end{equation}
For the spiral defect chaos states considered in this paper, the
velocity magnitude, $||\vec{u}||$, is about~$10$.  Thus, a simulation
at the Lewis number of, say, $\lew=10^{-3}$ will require $\Delta x
\sim 10^{-2}$ in order to satisfy Eq.~(\ref{eq:mesh-reynolds}).
However, current computational resources dictate that $\Delta x$ be
about $\gtapprox 10^{-1}$, and in fact, for the results quoted in this
paper, $\Delta x = 1/8$.  This problem is overcome by using a
filtering procedure developed by Fischer and Mullen~\cite{Fischer01},
and is described in Appendix~\ref{se:app}.  Using this filter and
maintaining $\Delta x =1/8$, a Lewis number as small as
$\lew=10^{-3}$ can be attained in a stable simulation.  The accuracy
of using the filter is also discussed in the Appendex.

\section{Results}
\label{se:res}
The transport equation, Eq.~(\ref{eq:tpt}), was integrated concurrently
with the Boussinesq equations, Eqs.~(\ref{eq:NS})--(\ref{eq:incom}),
for the following parameters: the Rayleigh number varied from the
onset of spiral defect chaos at $R \approx 3000$ to fully developed
spiral defect chaos at $R \approx 4000$, the Prandtl number
$\sigma=1$, and the Lewis number ranging from $\lew=10^{-3}$ to
$\lew=10^{-1}$.  The direct numerical simulations have been performed in
cylindrical three-dimensional cells of various aspect ratios.  In this
paper, data for an aspect ratio of $\Gamma=30$ will be reported.  The
initial condition used for the passive tracer concentration field is a
localized concentration at the center of the cell:
\begin{equation}
  \label{eq:psi_init}
\psi(x,y,z,t=0) = \exp\left[ -\frac{x^2+y^2+z^2}{6\Delta^2} \right],
\end{equation}
with $\Delta=1/4$ a small constant to ensure that the passive tracer
concentration is initially localized.  At time $t=0$, the temperature,
velocity, and pressure fields correspond to an asymptotic state of
spiral defect chaos, that is, one that has been evolved from random
thermal perturbations up to a time of $\mathcal{O}(\Gamma^2)$.  In
this paper, the focus will be on cells of large aspect ratio, $\Gamma
\ge 20$.  For these aspect ratios, the $z$-dependence of the passive
tracer concentration field was found to be essentially constant.  As
such, the $z$-dependence will be dropped in subsequent discussions and
the passive tracer concentration $\psi(x,y,t)$ will be considered as a
function of two-dimensional horizontal space and time.

\begin{figure*}
  \begin{center}
    \includegraphics[width=\widefigwidth]{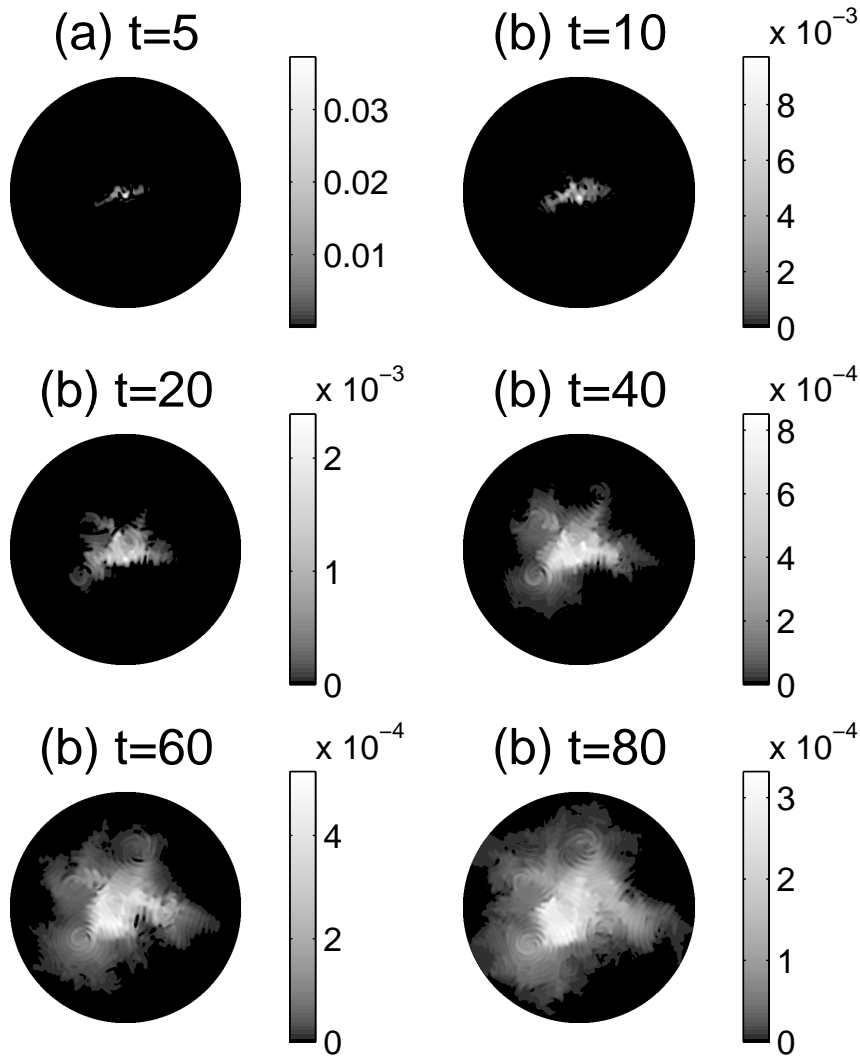}
  \end{center}
  \caption[Evolution of the passive tracer concentration field.]{
    Evolution of the passive tracer concentration field
    $\psi(x,y,z=0,t)$ for various times $t$, obtained by numerically
    solving Eq.~(\ref{eq:tpt}).  The Rayleigh number $R=3500$, the
    Prandtl number $\sigma=1$, and the Lewis number $\lew=10^{-2}$, and
    the cylindrical cell has the aspect ratio $\Gamma=30$.  The
    initial condition for $\psi$ at $t=0$ is given by
    Eq.~(\ref{eq:psi_init}).}
  \label{fig:tpt}
\end{figure*}

In Fig.~\ref{fig:tpt}, the evolution of the passive tracer
concentration field $\psi(x,y,t)$ at the mid-plane $z=0$ for the
parameters $R=3500$, $\sigma=1$, and $\lew=10^{-2}$ is shown for
various times $t$.  The passive tracer concentration spreads outward
with time in a non-uniform and non-axisymmetric way.  In
Sect.~\ref{se:tpt_stat}, this spreading is quantified globally by
studying the mean square displacement of the passive tracer
concentration field.  In Sect.~\ref{se:gaussian}, this spreading is
shown to be characterized by normal diffusion.  In
Sect.~\ref{se:tpt_ani}, the local dependence of the spreading on the
local wave number is discussed.

\subsection{Statistics of moments of passive tracer concentration}
\label{se:tpt_stat}

The spreading of the passive tracers can be quantified by its mean
square displacement $M_2(t)$, or the second moment, of the passive
tracer concentration field,
\begin{equation}
  \label{eq:variance}
  M_2(t) = \frac{\int_0^\Gamma \int_0^{2\pi} |\vec{x}- \langle
    \vec{x} \rangle(t)|^2 \psi(r,\theta,t)
    \,r\,dr\,d\theta}{\int_0^\Gamma \int_0^{2\pi}
    \psi(r,\theta,t)\,r\,dr\,d\theta}.
\end{equation}
Here, $\vec{x}=(r,\theta)$ is the polar coordinate with origin at the
center of the cell.  In practice, $M_2(t)$ is computed as the average
over different instances (typically three to five) of $\psi$ obtained
from different random initial conditions of spiral defect chaos, that
is, the velocity, temperature, and pressure fields used at $t=0$ are
different instances of fully developed spiral defect chaos.  The
quantity $\langle \vec{x} \rangle(t)$ is the instantaneous center of
mass of the tracer distribution,
\begin{equation}
  \label{eq:mean}
    \langle \vec{x} \rangle(t) = \frac{\int_0^\Gamma \int_0^{2\pi} \vec{x}
    \psi(r,\theta,t) \,r\,dr\,d\theta}{\int_0^\Gamma \int_0^{2\pi}
    \psi(r,\theta,t)\,r\,dr\,d\theta}.
\end{equation}

\begin{figure}
  \begin{center}
    \includegraphics[width=\figwidth]{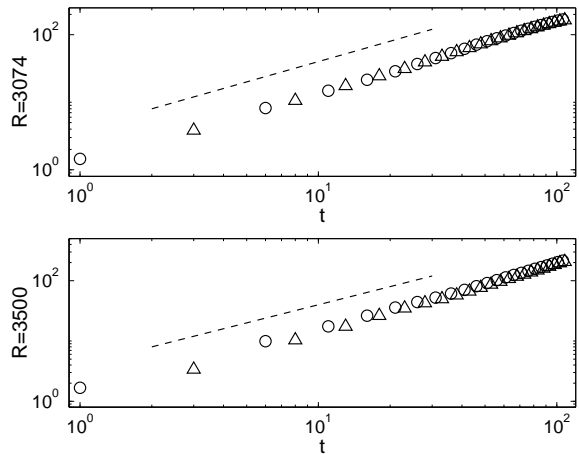}
  \end{center}
  \caption[The mean square displacement of the passive tracer concentration
  field.]{The mean square displacement $M_2(t)$ of the passive tracer
    concentration field for two different Rayleigh numbers $R=3074$
    (top) and $3500$ (bottom).  The Prandtl number is $\sigma=1$ in
    both cases.  The triangle and circle symbols denote data for the
    Lewis number $\lew=10^{-3}$ and $\lew=10^{-2}$, respectively.  The
    dashed lines have slopes of unity.  The exponents obtained from
    power law fits of the data are given in Table~\ref{tab:gamma}.}
  \label{fig:variance}
\end{figure}

In Fig.~\ref{fig:variance}, the mean square displacement $M_2(t)$ is
plotted for several different values of the Rayleigh number $R$ and
the Lewis numbers $\lew=10^{-3}$ and $\lew=10^{-2}$.  It is found
that, in all cases, the mean square displacement $M_2(t)$ is directly
proportional to the time $t$ to a very good approximation.  Least
squares fits of $M_2(t)$ to power laws $\sim t^\gamma$ yield exponents
$\gamma$ of approximately unity, as shown in Table~\ref{tab:gamma}.

\begin{table}
  \centering
  \begin{ruledtabular}
  \begin{tabular}{cccc}
        & $\lew$ & $10^{-2}$ & $10^{-3}$ \\ 
    $R$ & & & \\
    $3074$ & & 1.1 & 1.1 \\ 
    $3500$ & & 1.0 & 1.1 \\ 
    $4270$ & & 1.1 & 1.1
  \end{tabular}
  \end{ruledtabular}
  \caption{The exponent $\gamma$ computed from the fit of the mean
    square displacement $M_2(t)$ to a power law $\sim t^\gamma$ for
    several different values of the Rayleigh number $R$ and Lewis
    number $\lew$.  It is approximately unity in all instances.}
  \label{tab:gamma}
\end{table}

This implies that the spreading of the passive tracer concentration
field can be described by a normal diffusive process.  In other words,
the averaged passive tracer concentration, $\bar{\psi}(r,t)$, evolves
according to the one-dimensional normal diffusion equation,
\begin{equation}
  \label{eq:normal_dif}
  \partial_t \tilde{\psi}(r,t) = \leff \partial_{rr} \tilde{\psi}
\end{equation}
with $\leff$ an effective Lewis number that can be extracted from the
mean square displacement as
\begin{equation}
  \label{eq:leff}
  M_2(t) = 4 \leff t.
\end{equation}
Several values of the effective Lewis number for various values of the
parameters are tabulated in Table~\ref{tab:leff}.  

\begin{table}
  \centering
  \begin{ruledtabular}
  \begin{tabular}{ccccc}
    & $\lew$ & $10^{-3}$ & $10^{-2}$ & $10^{-1}$ \\  
    $R$ & & & & \\
    $3074$ & & 0.29 & 0.29 & 0.38 \\
    $3500$ & & 0.35 & 0.35 & 0.50 \\
    $4270$ & & 0.42 & 0.41 & 0.53
  \end{tabular}
  \end{ruledtabular}
  \caption{The effective Lewis number $\leff$ computed from
     Eq.~(\ref{eq:leff}) for various values of the Rayleigh number $R$
     and the Lewis number $\lew$.}
  \label{tab:leff}
\end{table}

\begin{figure}
  \begin{center}
    \includegraphics[width=\figwidth]{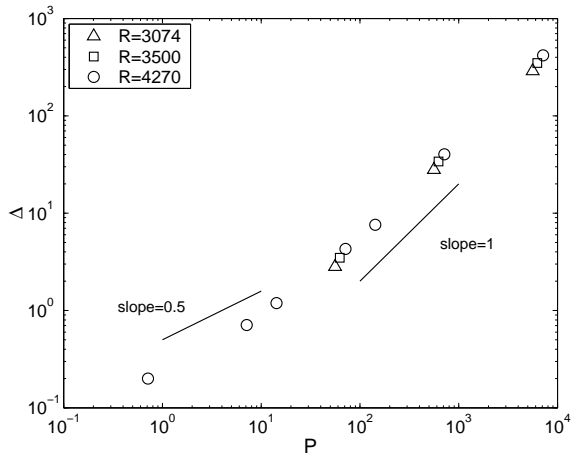}
  \end{center}
  \caption[The enhancement in Lewis number $\Delta$ vs. the P\'eclet
  number $\pec$.]{The dimensionless enhancement in molecular
    diffusivity $\Delta$ defined in Eq.~(\ref{eq:enh}) vs. the
    P\'eclet number $\pec$ for various Rayleigh numbers $R$ and the
    Prandtl number $\sigma=1$.  Note that, when the P\'eclet number
    approaches zero (that is, when the advection becomes negligible so
    that the transport equation, Eq.~(\ref{eq:tpt}), is the diffusion
    equation), the enhancement should approach zero as well.  Thus,
    the data points are expected to pass through the origin
    $(\pec=0,\Delta=0)$.}
  \label{fig:enh}
\end{figure}

The quantity
\begin{equation}
  \label{eq:enh}
  \Delta \equiv \frac{\leff-\lew}{\lew}
\end{equation}
is then a dimensionless measure of the enhancement in the molecular
diffusivity of the passive tracer concentration brought upon by the
advection of the spiral defect chaotic flow.  The goal of this paper
can then be phrased as the calculation of how this enhancement varies
as a function of both the properties of the advecting fluid (the
Rayleigh number $R$ and Prandtl number $\sigma$) and the property of
the passive tracer concentration [its Lewis number, or equivalently,
the P\'eclet number, $\pec$, c.f.~Eq.~(\ref{eq:peclet})],
\begin{equation}
  \Delta = \Delta( R, \sigma, \pec ).
\end{equation}
The results of this calculation are shown in Fig.~\ref{fig:enh}, which
depicts how the enhancement $\Delta$ varies vs.  the P\'eclet number
$\pec$ for various Rayleigh numbers $R$ and the Prandtl number
$\sigma=1$.

\begin{figure}[t]
  \begin{center}
    \includegraphics[width=\figwidth]{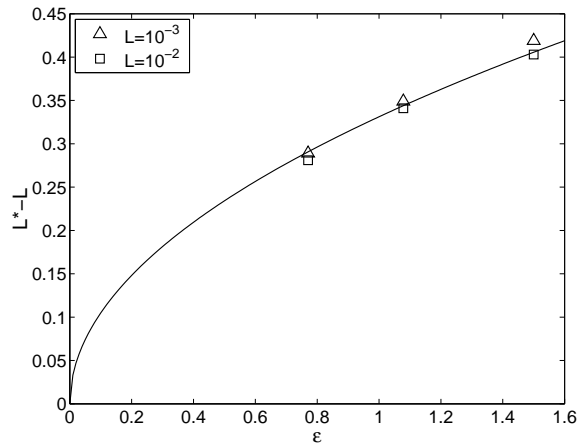}
  \end{center}
  \caption[The effective Lewis number vs. the reduced Rayleigh
  number.]{The effective Lewis number $\leff-\lew$ vs.  the reduced
    Rayleigh number $\epsilon$.  The solid line represents a power law
    with an exponent of $1/2$.}
  \label{fig:eff_Lew}
\end{figure}

It is found that the enhancement $\Delta$ follows two different
scaling regimes in the P\'eclet number $\pec$.  In the regime of large
P\'eclet numbers, $\pec \gtapprox 10^2$, the enhancement is found to
scale linearly with the P\'eclet number,
\begin{equation}
  \label{eq:tpt_adv}
  \Delta \propto \pec.
\end{equation}
This implies that the effective Lewis number also scales linearly with
the velocity magnitude of the flow, and is independent of the Lewis
number $\lew$,
\begin{equation}
  \label{eq:tpt_adv_2}
  \leff-\lew \propto ||\vec{u}||.
\end{equation}
[Eq.~(\ref{eq:tpt_adv_2}) is easily obtained by dividing
Eq.~(\ref{eq:tpt_adv}) by the Lewis number $\lew$.]  In addition, it
is of interest to see how the effective Lewis number relates to the
reduced Rayleigh number.  This relation is plotted in
Fig.~\ref{fig:eff_Lew}, and it exhibits a square-root dependence,
\begin{equation}
  \label{eq:tpt_adv_3}
  \leff-\lew \propto \epsilon^{1/2}.
\end{equation}
Thus, Eqs.~(\ref{eq:tpt_adv_2})~and~(\ref{eq:tpt_adv_3}) together suggest that 
the characteristic velocity scale of spiral defect chaos scales with
the reduced Rayleigh number as
\begin{equation}
  ||\vec{u}|| \propto \epsilon^{1/2}.
\end{equation}

On the other hand, in the regime of small P\'eclet numbers, $\pec
\ltapprox 10^2$, the enahancement is found to scale with the P\'eclet
number as
\begin{equation}
  \label{eq:tpt_shraiman}
 \Delta \propto \pec^{1/2}.
\end{equation}
(Note that there are insufficient data to conclude whether the
crossover between the two regimes is a continuous and gradual one or a
discontinuous and sharp one.)  This square root dependence of the
enhancement on the P\'eclet number is similar to the result obtained
experimentally~\cite{Solomon88} and calculated
theoretically~\cite{Rosenbluth87,Shraiman87} in the spreading of
passive tracers in time-independent convection flows comprising
straight parallel rolls.  In this case, the enhancement can be
attributed to the expulsion of the gradient of the passive tracer
concentration from regions of closed stream lines~\cite{Shraiman87}.
Near a separatrix between two sets of closed stream lines, the only
transport of the passive tracers from one roll to the next comes from
the random walks of the passive tracers that lie within a thin layer
of width $d$ of the roll boundary (the roll itself is of unit width in
the dimensionless unit system adopted in this paper).  Thus, a
fraction $d$ of passive tracers contribute to an increase, $d\lew$, in
the effective Lewis number of the diffusion.  The width $d$ can be
estimated from dimensional analysis~\cite{Bohr98} to be $d^2 \sim
\pec^{-1}$.  Combining these estimates leads immediately to
Eq.~(\ref{eq:tpt_shraiman}).  Thus, the result above suggests that the
gradient expulsion mechanism near closed streamlines, although
strictly derived in a time-independent convection flows comprising
straight parallel rolls, may be a universal mechanism at sufficiently
small P\'eclet numbers independent of the structure of the underlying
flow field.

In Sect.~\ref{se:tpt_ani}, the origin of these two distinct regimes is
discussed in terms of the dependence on the local wave number of the
convection rolls.  But before concluding this section, some details
are presented in the way the least squares fits were performed.  First,
data from early times are ignored because of the presence of
transients.  One such transient effect could be that, at very early
times prior to the turnover time scale $\tau_c \sim ||\vec{u}||^{-1}
\sim \mathcal{O}(10^{-1})$, the passive tracers ``feel'' that they are
being transported by a constant velocity field, and so will exhibit
ballistic behavior with $\gamma = 2$.  There is then a crossover time
in which $\gamma$ decreases to unity, and this regime is to be ignored
too.  Second, data from late times are also ignored because of finite
size effects.  The time at which finite size effects become important
is chosen as the time at which the exponent $\gamma$, obtained from
the logarithmic derivative
\begin{equation}
  \gamma(t) = \frac{d \log[M_2(t)]}{d \log(t)},
\end{equation}
deviates from approximately unity for a purely diffusive process with
without advection (that is, whose diffusivity is chosen to match that
of effective diffusivity of the above process).


\subsection{Normal diffusion vs. anomalous diffusion}
\label{se:gaussian}

In this section, results are discussed from two other tests that show
that the spreading process is indeed governed by normal diffusion, and
not anomalous diffusion.  Diffusion is said to be anomalous when the
mean square displacement is not proportional to time, that is, when
$M_2(t) \propto t^\gamma$ with the exponent $\gamma \ne 1$.  Anomalous
diffusion has been observed in the transport of passive tracers in
cellular Taylor-Couette flow in a rotating
annulus~\cite{Solomon93,Solomon94}, and in various other geophysical
turbulent flows arising from the presence of L\'evy
trajectories~\cite{Shlesinger95}.  However, results from this section
show no evidence of anomalous diffusion in transport in spiral defect
chaos for the range of Lewis numbers investigated.

\begin{figure}
  \begin{center}
    \includegraphics[width=\figwidth]{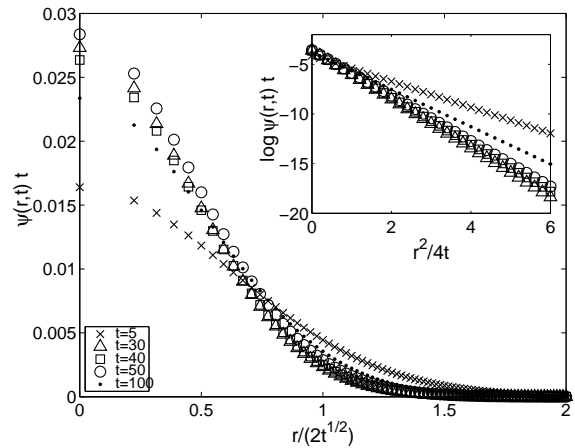}
  \end{center}
  \caption[The scaled passive tracer concentration
  $t\tilde{\psi}(r,t)$ vs. the scaled distance squared $r^2/4t$.]{ The
    scaled passive tracer concentration $t\tilde{\psi}(r,t)$ vs. the
    scaled distance $\sqrt{r^2/4t}$ and scaled distance squared
    $r^2/4t$ (inset).  The various symbols denote different values of
    the times $t$.  The data for times $t=30$, $t=40$, and $t=70$
    (triangles, squares, and circles, respectively) collapse onto the
    same curve, suggesting the validity of the Gaussian form in this
    time range.}
  \label{fig:tpt_gaussian}
\end{figure}

First, if the passive tracer concentration is spreading by normal
diffusion and so obeys Eq.~(\ref{eq:normal_dif}), then it can be
expressed in the form of a Gaussian,
\begin{equation}
  \tilde{\psi}(r,t) \sim \frac{1}{t} \exp\left( \frac{r^2}{4\leff t}
  \right),
\end{equation}
and a plot of the logarithm of the scaled passive tracer concentration
$\log [ t \tilde{\psi}(r,t)]$ vs.~the scaled distance  $\sqrt{r^2/4t}$
for different times $t$ will all collapse onto the same curve.
This is indeed the case, as shown in Fig.~\ref{fig:tpt_gaussian},
which shows the data at times $t=30$, $t=40$, and $t=70$ (triangles,
squares, and circles, respectively) collapsing onto the same straight
line.  However, data from an earlier time $t=5$ (crosses) do not
collapse onto the same straight line, presumably because of the
presence of transient effects.  Similarly, data from a later time
$t=100$ (dots) do not collapse onto the same straight line, because of
the presence of finite size effects.

\begin{figure}
  \begin{center}
    \includegraphics[width=\figwidth]{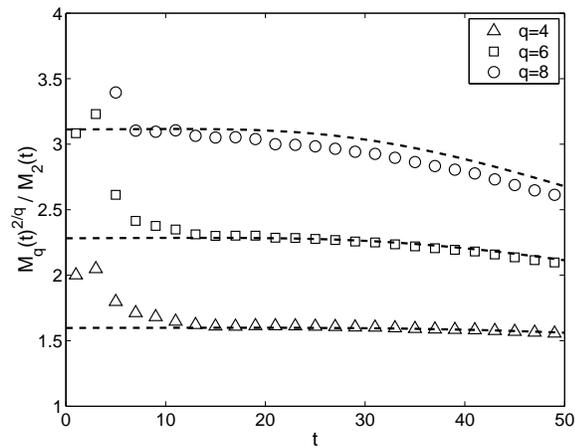}
  \end{center}
  \caption[Scaled higher-order moments of the passive tracer concentration.]{
    The ratio of the scaled higher-order moments to the second-order
    moment, $M_q(t)^{2/q}/M_2(t)$ , of the passive tracer
    concentration vs.  time $t$ for the parameters $R=3500$,
    $\sigma=1$, and $\lew=10^{-2}$, and for a purely diffusive process
    (dashed lines).  From Eq.~(\ref{eq:scaled}), this ratio is $1.73$
    when $q=4$, $2.47$ when $q=6$, and $3.20$ when $q=8$.  As $q$
    increases, the range of time for which the scaled moment stays
    constant in time decreases, because of finite size effects.}
  \label{fig:mult}
\end{figure}

Second, possible deviations from the Gaussian behavior of the passive
tracer concentration can be checked by looking at the higher-order
moments,
\begin{equation}
  M_q(t) =  \frac{\int_0^\Gamma \int_0^{2\pi} |\vec{x}-
    \langle \vec{x}\rangle(t)|^q \psi(r,\theta,t) \,r\,dr\,d\theta}{\int_0^\Gamma
    \int_0^{2\pi} \psi(r,\theta,t)\,r\,dr\,d\theta}.
\end{equation}
For normal diffusion, the higher-order moments scale like
\begin{equation}
  M_q(t) \propto t^{q/2},
\end{equation}
and the ratio of this higher-order moment scaled to the second-order
moment can be calculated to be
\begin{equation}
  \label{eq:scaled}
  \frac{M_q(t)^{2/q}}{M_2(t)} = 1\cdot 3 \cdots (q-1) = (q-1)!!
\end{equation}
which is a constant in time.  In Fig.~\ref{fig:mult}, this scaled
ratio is plotted for $q=4$,~$6$,~and~$8$ as functions of time when the
Rayleigh number $R=3500$, the Prandtl number $\sigma=1$, and the Lewis
number $\lew=10^{-2}$.  The dashed lines show the corresponding
quantity for a purely diffusive process with diffusivity chosen to
match the former's effective diffusivity.  (Because of finite size
effects, the scaled higher-order moments unfortunately have only a
small range for which they are constant in time.  For example, at
$q=8$, this range is only $5 \ltapprox t \ltapprox 20$.)  The
agreement of the two sets of data shows that, apart from finite size
effects, there are no discernible deviations from Gaussian form for
the passive tracer concentrations.

Thus, both observations above suggest that the spreading of the
passive tracer concentration is governed by normal diffusion.

\subsection{Wave number dependence of the passive tracer transport}
\label{se:tpt_ani}

\begin{figure}
  \begin{center}
    \includegraphics[width=\figwidth]{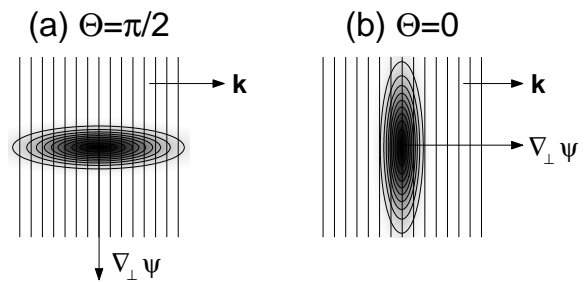}
  \end{center}
  \caption[Illustrations showing the definition of the horizontal spreading
  orientation.]{Illustrations showing the definition of the horizontal
    spreading orientation, $\Theta$, at \emph{(a)} $\Theta=\pi/2$
    corresponding to spreading in the direction of the local wave
    vector $\vec{k}$, and \emph{(b)} $\Theta=0$ corresponding to
    spreading in the direction orthogonal to $\vec{k}$.}
  \label{fig:tpt_orient}
\end{figure}

\begin{figure}
  \begin{center}
    \includegraphics[width=\figwidth]{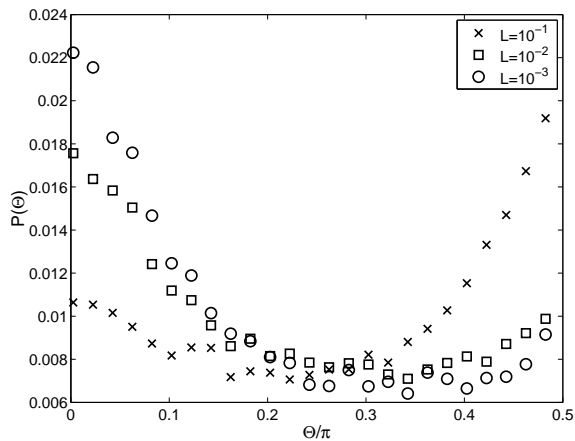}
  \end{center}
  \caption[Distribution of horizontal spreading orientations
  $P(\Theta)$ for various Lewis numbers.]{Distribution of horizontal
  spreading orientations $P(\Theta)$ for the Rayleigh number $R=3500$,
  the Prandtl number $\sigma=1$ and the Lewis number ranging from
  $\lew=10^{-3}$ to $\lew=10^{-1}$.}
  \label{fig:dot_angle}
\end{figure}

In this section, the existence of two different scaling regimes for
the dimensionless enhancement in molecular diffusivity, namely at
large P\'eclet numbers given by Eq.~(\ref{eq:tpt_adv}) and at small
P\'eclet numbers given by Eq.~(\ref{eq:tpt_shraiman}), is investigated
in terms of the local wave number dependence of the passive tracer
concentration.  To make this discussion more quantitative, first a
quantity called the horizontal spreading orientation, $\Theta(x,y)$,
is calculated at every location in the cell,
\begin{equation}
  \cos(\Theta) = \frac{\vec{\nabla}_\perp \psi \dotprod
  \vec{k}}{|\vec{\nabla}_\perp \psi| | \vec{k}|}.
\end{equation}
The subscript $\perp$ denotes the horizontal coordinates $(x,y)$ and
$\vec{k}(x,y)$ is the local wave vector at location $(x,y)$ in the
planform~\cite{Egolf98}.  If the passive tracer concentration spreads
in the direction of the local wave vector $\vec{k}$, then, as
illustrated in Fig.~\ref{fig:tpt_orient}(a), the gradient
$\vec{\nabla}_\perp \psi$ will be orthogonal to $\vec{k}$, resulting
in the local horizontal spreading orientation acquiring the value of
$\Theta=\pi/2$.  On the other hand, if the passive tracer
concentration spreads in the direction orthogonal to $\vec{k}$, then,
as illustrated in Fig.~\ref{fig:tpt_orient}(b), the local horizontal
spreading orientation will be $\Theta=0$.
For a particular passive tracer concentration, the horizontal
spreading orientation can be computed locally at every point in the
mid-plane of the convection cell and then were sorted into bins to
create a histogram.  In Fig.~\ref{fig:dot_angle}, such a distribution
of the horizontal spreading orientation, $P(\Theta)$, is plotted for
several values of the Lewis number ranging from $\lew=10^{-3}$ to
$\lew=10^{-1}$, the Rayleigh number $R=3500$, and the Prandtl number
$\sigma=1$ at time $t=50$.  Consider first the distribution for the
relatively large Lewis number of $\lew=10^{-1}$ (denoted by crosses).
This distribution has a peak at $\Theta=\pi/2$.  This suggests that
the spreading of the passive tracer concentration has the highest
probability to be in the direction of the wave vector $\vec{k}$ [that
is, as illustrated in the scenario of Fig.~\ref{fig:tpt_orient}(a)].
This is consistent with the gradient expulsion mechanism near closed
streamlines described earlier in Sect.~\ref{se:tpt_stat}.  However, at
the smaller Lewis numbers of, say, $\lew=10^{-3}$, the distribution of
the horizontal spreading orientation is visibly different.  The
distribution now has a peak at $\Theta=0$.  In other words, the
spreading of the passive tracer concentration has the highest
probability in the direction orthogonal to the wave vector $\vec{k}$
[that is, as illustrated in scenario of Fig.~\ref{fig:tpt_orient}(b)].
This shows that two different scaling regimes for the dimensionless
enhancement in molecular diffusivity are associated with two different
transport mechanisms.  At large Lewis numbers (or equivalently, small
P\'eclet numbers), the transport is along the direction of the wave
vector $\vec{k}$ by the gradient expulsion mechanism.  At small Lewis
numbers (or equivalently, large P\'eclet numbers), the transport is
orthogonal to the wave vector $\vec{k}$, presumably by advection by
the disordered flow.

\begin{figure}
  \begin{center}
    \includegraphics[width=\figwidth]{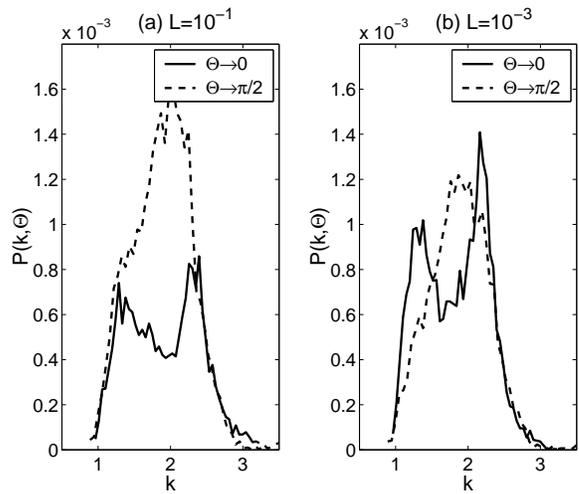}
  \end{center}
  \caption[Distribution of wave numbers $P(k,\Theta)$ for which
  longitudinal and lateral spreadings occur.]{Distribution of wave
    numbers $P(k,\Theta)$ for which spreading occurs orthogonally to
    the local wave vector $\vec{k}$ (solid lines) and in the direction
    of $\vec{k}$ (dashed lines), for \emph{(a)} large Lewis number
    $\lew=10^{-1}$ and \emph{(b)} small Lewis number $\lew=10^{-3}$ at
    the Rayleigh number $R=3500$ and the Prandtl number $\sigma=1$.}
  \label{fig:dot_angle_k}
\end{figure}

To determine why there is transport orthogonal to the wave vector
$\vec{k}$ at small Lewis numbers (or equivalently, large P\'eclet
numbers), the following calculation was performed.  The local wave
numbers that correspond to locations that exhibit spreading in the
direction orthogonal to the wave vector $\vec{k}$ (that is, $0 \le
\Theta \le \eta$ where $\eta = 0.01$ is a small constant) were
compared with those locations that exhibit spreading in the direction
of the wave vector $\vec{k}$ (that is, $\pi/2 - \eta \le \Theta \le
\pi/2$).  In Fig.~\ref{fig:dot_angle_k}, the distribution of wave
numbers $P(k,\Theta)$ for which the spreading occurs orthogonally to
$\vec{k}$ (solid lines) is plotted together with the distribution for
which spreading is along $\vec{k}$ (dashed lines) for a large Lewis
number case [Fig.~\ref{fig:dot_angle_k}(a)] and a small Lewis number
case [Fig.~\ref{fig:dot_angle_k}(b)].  The key point to observe is
that, in the small Lewis number case, there is a higher probability
that spreading occurs orthogonal to the wave vector ($\Theta
\rightarrow 0$, solid line) than along the wave vector ($\Theta
\rightarrow \pi/2$, dashed line) for wave numbers $k \approx 1.5$ and
$k \approx 2.5$.  These wave numbers, far away from the mean wave
number, correspond to to the occurrence of defects such as spiral
cores, target cores, dislocations, etc.  This suggests that the reason
why gradient expulsion ceases to be valid at large P\'eclet numbers is
due to the enhanced transport of the passive tracers orthogonal to the
local wave vectors by the defects in the pattern.  However, the
presence of defects is not sufficient to overcome the gradient
expulsion mechanism at small P\'eclet numbers.

\section{Conclusions}
\label{se:con}
In this paper, the spreading of a passive tracer concentration in a
\rbc\ flow exhibiting spiral defect chaos is studied for the first
time.  All previous studies have dealt with time-independent or
oscillatory flows.  In the presence of advection by spiral defect
chaos, we find that the spreading continues to be characterized by
normal diffusion.  The enhancement follows two regimes.  When the
P\'eclet number is large (that is, when the molecular diffusivity of
the tracer is small), the enhancement is proportional to the P\'eclet
number.  This means that in the limit of large P\'eclet number the
effective diffusivity is independent of the molecular diffusivity, and
is proportional to the strength of the advection velocity field.  When
the P\'eclet numbers is small, the enhancement is proportional to the
square root of the P\'eclet number.  These results are explained in
terms of the dependence of the transport on the local wave numbers.
It is found that tracers with small P\'eclet numbers follow the
gradient expulsion mechanism described previously in time-independent
flows \cite{Shraiman87} which predicts the square-root dependence.
However, when the P\'eclet number becomes large, defects in the flow
field became important and lead to enhanced transport orthogonal to the
local wave vectors.

\begin{acknowledgments}
  
  This work was supported by the Engineering Research Program of the
  Office of Basic Energy Sciences at the Department of Energy, Grants
  DE-FG03-98ER14891 and DE-FG02-98ER14892.  We acknowledge the Caltech
  Center for Advanced Computing Research and the North Carolina
  Supercomputing Center.  We thank Tony Leonard and Dan Meiron for
  useful discussions, as well as Janet Scheel for her generous help
  with carrying out the computations.

\end{acknowledgments}

\appendix
\section{Numerical details}
\label{se:app}

As mentioned in Sect.~\ref{se:dns}, for small Lewis numbers $\lew \ll
1$, a difficulty associated with integrating the transport equation,
Eq.~(\ref{eq:tpt}), is that the spatial resolution $\Delta x$ must be
small.  A larger $\Delta x$ can be used by employing a simple
filtering procedure developed by Fischer and Mullen~\cite{Fischer01}.
At the end of each time step, a filter is applied on an
element-by-element basis to the passive tracer concentration field
$\psi$.  In one dimension, the filtered field can be written as
\begin{equation}
  \label{eq:filter}
  F(\psi;\alpha) = \alpha \Pi_{N-1}(\psi) + (1-\alpha) \psi,
\end{equation}
where the operator $\Pi_{N-1}$ first interpolates $\psi$ onto the mesh
points for a polynomial of degree $N-1$ determined by the mesh spacing
and then interpolates the result back onto the mesh points for a
polynomial for degree $N$.  In higher dimensions, the tensor product
form of Eq.~(\ref{eq:filter}) is used.  Typical values of $\alpha$
used were in the range $0.05 \le \alpha \le 0.2$.  This filtering
procedure preserves inter-element continuity and spectral accuracy.
Using this filter and maintaining $\Delta x =1/8$, a stable Lewis
number of up to $\lew=10^{-3}$ could be attained.

To verify that the filter allows the diffusion to be sufficiently and
accurately resolved at small Lewis numbers, two sets of checks were
performed.

\begin{figure}
  \begin{center}
    \includegraphics[width=\figwidth]{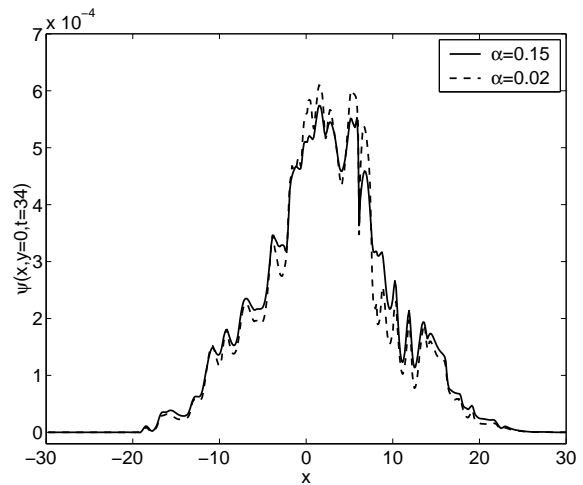}
  \end{center}
  \caption{Snapshot of the passive tracer concentration field
    $\psi(x,y=0)$ for $\lew=10^{-2}$ at time $t=34$ for two different
    values of the filtering parameter $\alpha=0.15$ (solid line) and
    $\alpha=0.02$ (dashed line).}
  \label{fig:verify_alpha}
\end{figure}

First, the passive tracer concentration field $\psi$ is inspected for
various values of $\alpha$.  In Fig.~\ref{fig:verify_alpha}, a
snapshot of the passive tracer concentration field $\psi$ for $\lew =
10^{-2}$ at a time $t=34$ is shown for two values of $\alpha=0.15$
(value used throughout in this paper) and $\alpha=0.02$.  It can be
seen that, although differences exist between the peak values of the
filtered and the solution with $\alpha=0.02$, we can see that the
extent and reach of the solution is largely unaffected it terms of the
range of $x$ it covers.  Thus, when computing, say, the second moment
of the passive tracer concentration field [Eq.~(\ref{eq:variance})],
the results will be largely independent of $\alpha$.  This justifies
that it is safe to use a filter parameter of as high as $\alpha=0.15$
when computing statistics of the passive scalar concentration field.

\begin{figure}
  \begin{center}
    \includegraphics[width=\figwidth]{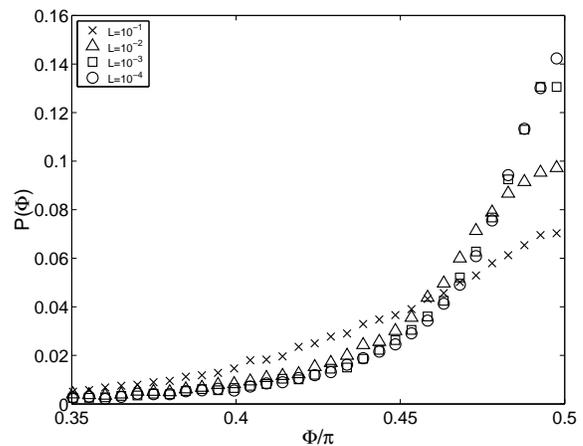}
  \end{center}
  \caption[Distribution of local advection orientations for various
  Lewis numbers.]{Distribution of local advection orientations for
  various Lewis numbers ranging from $\lew=10^{-4}$ to $\lew=10^{-1}$.}
  \label{fig:dot_vel_angle}
\end{figure}

Second, a local advection orientation, $\Phi(x,y)$, is
defined:
\begin{equation}
  \cos(\Phi) = \frac{\vec{\nabla} \psi \dotprod
  \vec{u}}{|\vec{\nabla} \psi| | \vec{u}|},
\end{equation}
with $\vec{u}$ the velocity field.  If the local passive tracer
concentration is being advected by the local velocity and diffusion is
not being sufficiently resolved, then the gradient of the former will
be orthogonal to the local velocity, and consequently, $\Phi=\pi/2$.
On the other hand, if the local passive tracer concentration exhibits
diffusion, then it will change in a direction perpendicular to the
local velocity, yielding $\Phi=0$.  For small Lewis numbers where the
effects of advection dominate over the effects of molecular diffusion,
the distribution of the local advection orientation, $P(\Phi)$, over
the mid-plane of the cell, should exhibit a strong peak at
$\Phi=\pi/2$.  This peak will then broaden as the Lewis number is
increased, since the effects of diffusion cause the passive tracer
concentration to spread out at all orientations relative to the local
velocity.  The presence of this broadening in the distribution of the
local advection orientation is then an indication that molecular
diffusion has been sufficiently resolved.  The distributions $P(\Phi)$
for the various Lewis numbers ranging from $\lew=10^{-4}$ to
$\lew=10^{-1}$ are plotted in Fig.~\ref{fig:dot_vel_angle}.
The distribution for $\lew=10^{-2}$ is distinctly different from that for
$\lew=10^{-3}$, providing evidence that the molecular diffusion at
$\lew=10^{-3}$ has been stably resolved, that is, that the chosen grid
spacing $\Delta x$ is sufficiently small for the simulation to be
accurate.  However, the relative similarity in the distributions for
$\lew=10^{-3}$ and $\lew=10^{-4}$ suggests that diffusion for the latter
case may not have been sufficiently resolved.  Consequently, the
smallest allowed Lewis number is set at $\lew=10^{-3}$.

\bibliographystyle{apsrev}
\bibliography{books_theses,cfd,pattern,turbulence,misc}

\begin{thebibliography}{30}
\expandafter\ifx\csname natexlab\endcsname\relax\def\natexlab#1{#1}\fi
\expandafter\ifx\csname bibnamefont\endcsname\relax
  \def\bibnamefont#1{#1}\fi
\expandafter\ifx\csname bibfnamefont\endcsname\relax
  \def\bibfnamefont#1{#1}\fi
\expandafter\ifx\csname citenamefont\endcsname\relax
  \def\citenamefont#1{#1}\fi
\expandafter\ifx\csname url\endcsname\relax
  \def\url#1{\texttt{#1}}\fi
\expandafter\ifx\csname urlprefix\endcsname\relax\def\urlprefix{URL }\fi
\providecommand{\bibinfo}[2]{#2}
\providecommand{\eprint}[2][]{\url{#2}}

\bibitem[{\citenamefont{Cross and Hohenberg}(1994)}]{Cross94}
\bibinfo{author}{\bibfnamefont{M.~C.} \bibnamefont{Cross}} \bibnamefont{and}
  \bibinfo{author}{\bibfnamefont{P.~C.} \bibnamefont{Hohenberg}},
  \bibinfo{journal}{Science}
  \textbf{\bibinfo{volume}{263}}(\bibinfo{number}{5153}), \bibinfo{pages}{1569}
  (\bibinfo{year}{1994}).

\bibitem[{\citenamefont{Gollub and Cross}(2000)}]{Gollub00}
\bibinfo{author}{\bibfnamefont{J.~P.} \bibnamefont{Gollub}} \bibnamefont{and}
  \bibinfo{author}{\bibfnamefont{M.~C.} \bibnamefont{Cross}},
  \bibinfo{journal}{Nature}
  \textbf{\bibinfo{volume}{404}}(\bibinfo{number}{6779}), \bibinfo{pages}{710}
  (\bibinfo{year}{2000}).

\bibitem[{\citenamefont{Hohenberg and Shraiman}(1989)}]{Hohenberg89}
\bibinfo{author}{\bibfnamefont{P.~C.} \bibnamefont{Hohenberg}}
  \bibnamefont{and} \bibinfo{author}{\bibfnamefont{B.~I.}
  \bibnamefont{Shraiman}}, \bibinfo{journal}{Physica D}
  \textbf{\bibinfo{volume}{37}}, \bibinfo{pages}{109} (\bibinfo{year}{1989}).

\bibitem[{\citenamefont{Solomon and Gollub}(1988{\natexlab{a}})}]{Solomon88}
\bibinfo{author}{\bibfnamefont{T.~H.} \bibnamefont{Solomon}} \bibnamefont{and}
  \bibinfo{author}{\bibfnamefont{J.~P.} \bibnamefont{Gollub}},
  \bibinfo{journal}{Physics of Fluids}
  \textbf{\bibinfo{volume}{31}}(\bibinfo{number}{6}), \bibinfo{pages}{1372}
  (\bibinfo{year}{1988}{\natexlab{a}}).

\bibitem[{\citenamefont{Rosenbluth et~al.}(1987)\citenamefont{Rosenbluth, Berk,
  Doxas, and Horton}}]{Rosenbluth87}
\bibinfo{author}{\bibfnamefont{M.~N.} \bibnamefont{Rosenbluth}},
  \bibinfo{author}{\bibfnamefont{H.~L.} \bibnamefont{Berk}},
  \bibinfo{author}{\bibfnamefont{I.}~\bibnamefont{Doxas}}, \bibnamefont{and}
  \bibinfo{author}{\bibfnamefont{W.}~\bibnamefont{Horton}},
  \bibinfo{journal}{Physics of Fluids} \textbf{\bibinfo{volume}{30}},
  \bibinfo{pages}{2636} (\bibinfo{year}{1987}).

\bibitem[{\citenamefont{Shraiman}(1987)}]{Shraiman87}
\bibinfo{author}{\bibfnamefont{B.~I.} \bibnamefont{Shraiman}},
  \bibinfo{journal}{Physical Review A}
  \textbf{\bibinfo{volume}{36}}(\bibinfo{number}{1}), \bibinfo{pages}{261}
  (\bibinfo{year}{1987}).

\bibitem[{\citenamefont{McCarty and Horsthemke}(1988)}]{McCarty88}
\bibinfo{author}{\bibfnamefont{P.}~\bibnamefont{McCarty}} \bibnamefont{and}
  \bibinfo{author}{\bibfnamefont{W.}~\bibnamefont{Horsthemke}},
  \bibinfo{journal}{Physical Review A} \textbf{\bibinfo{volume}{37}},
  \bibinfo{pages}{2112} (\bibinfo{year}{1988}).

\bibitem[{\citenamefont{McLaughlin et~al.}(1985)\citenamefont{McLaughlin,
  Papanicolaou, and Pironneau}}]{McLaughlin85}
\bibinfo{author}{\bibfnamefont{D.~W.} \bibnamefont{McLaughlin}},
  \bibinfo{author}{\bibfnamefont{G.~C.} \bibnamefont{Papanicolaou}},
  \bibnamefont{and} \bibinfo{author}{\bibfnamefont{O.~R.}
  \bibnamefont{Pironneau}}, \bibinfo{journal}{SIAM Journal on Applied
  Mathematics} \textbf{\bibinfo{volume}{45}}, \bibinfo{pages}{780}
  (\bibinfo{year}{1985}).

\bibitem[{\citenamefont{Busse}(1978)}]{Busse78}
\bibinfo{author}{\bibfnamefont{F.~H.} \bibnamefont{Busse}},
  \bibinfo{journal}{Reports on Progress in Physics}
  \textbf{\bibinfo{volume}{41}}, \bibinfo{pages}{1929} (\bibinfo{year}{1978}).

\bibitem[{\citenamefont{Solomon and Gollub}(1988{\natexlab{b}})}]{Solomon88b}
\bibinfo{author}{\bibfnamefont{T.~H.} \bibnamefont{Solomon}} \bibnamefont{and}
  \bibinfo{author}{\bibfnamefont{J.~P.} \bibnamefont{Gollub}},
  \bibinfo{journal}{Physical Review A} \textbf{\bibinfo{volume}{38}},
  \bibinfo{pages}{6280} (\bibinfo{year}{1988}{\natexlab{b}}).

\bibitem[{\citenamefont{Camassa and Wiggins}(1991{\natexlab{a}})}]{Camassa91}
\bibinfo{author}{\bibfnamefont{R.}~\bibnamefont{Camassa}} \bibnamefont{and}
  \bibinfo{author}{\bibfnamefont{S.}~\bibnamefont{Wiggins}},
  \bibinfo{journal}{Physical Review A}
  \textbf{\bibinfo{volume}{43}}(\bibinfo{number}{2}), \bibinfo{pages}{774}
  (\bibinfo{year}{1991}{\natexlab{a}}).

\bibitem[{\citenamefont{Camassa and Wiggins}(1991{\natexlab{b}})}]{Camassa91b}
\bibinfo{author}{\bibfnamefont{R.}~\bibnamefont{Camassa}} \bibnamefont{and}
  \bibinfo{author}{\bibfnamefont{S.}~\bibnamefont{Wiggins}},
  \bibinfo{journal}{Physica D} \textbf{\bibinfo{volume}{51}},
  \bibinfo{pages}{472} (\bibinfo{year}{1991}{\natexlab{b}}).

\bibitem[{\citenamefont{Ramshankar et~al.}(1990)\citenamefont{Ramshankar,
  Berlin, and Gollub}}]{Ramshankar90}
\bibinfo{author}{\bibfnamefont{R.}~\bibnamefont{Ramshankar}},
  \bibinfo{author}{\bibfnamefont{D.}~\bibnamefont{Berlin}}, \bibnamefont{and}
  \bibinfo{author}{\bibfnamefont{J.~P.} \bibnamefont{Gollub}},
  \bibinfo{journal}{Physics of Fluids A}
  \textbf{\bibinfo{volume}{2}}(\bibinfo{number}{11}), \bibinfo{pages}{1955}
  (\bibinfo{year}{1990}).

\bibitem[{\citenamefont{Ramshankar and Gollub}(1991)}]{Ramshankar91}
\bibinfo{author}{\bibfnamefont{R.}~\bibnamefont{Ramshankar}} \bibnamefont{and}
  \bibinfo{author}{\bibfnamefont{J.~P.} \bibnamefont{Gollub}},
  \bibinfo{journal}{Physics of Fluids A}
  \textbf{\bibinfo{volume}{3}}(\bibinfo{number}{5}), \bibinfo{pages}{1344}
  (\bibinfo{year}{1991}).

\bibitem[{\citenamefont{Solomon et~al.}(1993)\citenamefont{Solomon, Weeks, and
  Swinney}}]{Solomon93}
\bibinfo{author}{\bibfnamefont{T.~H.} \bibnamefont{Solomon}},
  \bibinfo{author}{\bibfnamefont{E.~R.} \bibnamefont{Weeks}}, \bibnamefont{and}
  \bibinfo{author}{\bibfnamefont{H.~L.} \bibnamefont{Swinney}},
  \bibinfo{journal}{Physical Review Letters}
  \textbf{\bibinfo{volume}{71}}(\bibinfo{number}{24}), \bibinfo{pages}{3975}
  (\bibinfo{year}{1993}).

\bibitem[{\citenamefont{Solomon et~al.}(1994)\citenamefont{Solomon, Weeks, and
  Swinney}}]{Solomon94}
\bibinfo{author}{\bibfnamefont{T.~H.} \bibnamefont{Solomon}},
  \bibinfo{author}{\bibfnamefont{E.~R.} \bibnamefont{Weeks}}, \bibnamefont{and}
  \bibinfo{author}{\bibfnamefont{H.~L.} \bibnamefont{Swinney}},
  \bibinfo{journal}{Physica D} \textbf{\bibinfo{volume}{76}},
  \bibinfo{pages}{70} (\bibinfo{year}{1994}).

\bibitem[{\citenamefont{Cross and Hohenberg}(1993)}]{Cross93}
\bibinfo{author}{\bibfnamefont{M.~C.} \bibnamefont{Cross}} \bibnamefont{and}
  \bibinfo{author}{\bibfnamefont{P.~C.} \bibnamefont{Hohenberg}},
  \bibinfo{journal}{Review of Modern Physics}
  \textbf{\bibinfo{volume}{65}}(\bibinfo{number}{3}), \bibinfo{pages}{851}
  (\bibinfo{year}{1993}).

\bibitem[{\citenamefont{Aref}(1983)}]{Aref83}
\bibinfo{author}{\bibfnamefont{H.}~\bibnamefont{Aref}},
  \bibinfo{journal}{Journal of Fluid Mechanics} \textbf{\bibinfo{volume}{143}},
  \bibinfo{pages}{1} (\bibinfo{year}{1983}).

\bibitem[{\citenamefont{Fischer}(1997)}]{Fischer97}
\bibinfo{author}{\bibfnamefont{P.~F.} \bibnamefont{Fischer}},
  \bibinfo{journal}{Journal of Computational Physics}
  \textbf{\bibinfo{volume}{133}}(\bibinfo{number}{1}), \bibinfo{pages}{84}
  (\bibinfo{year}{1997}).

\bibitem[{\citenamefont{Chiam et~al.}(2003)\citenamefont{Chiam, Cross, and
  Greenside}}]{Chiam03}
\bibinfo{author}{\bibfnamefont{K.-H.} \bibnamefont{Chiam}},
  \bibinfo{author}{\bibfnamefont{M.~C.} \bibnamefont{Cross}}, \bibnamefont{and}
  \bibinfo{author}{\bibfnamefont{H.~S.} \bibnamefont{Greenside}},
  \bibinfo{journal}{Physical Review E} \textbf{\bibinfo{volume}{67}},
  \bibinfo{pages}{056206} (\bibinfo{year}{2003}).

\bibitem[{\citenamefont{Paul et~al.}(2002)\citenamefont{Paul, Cross, and
  Fischer}}]{Paul02}
\bibinfo{author}{\bibfnamefont{M.~R.} \bibnamefont{Paul}},
  \bibinfo{author}{\bibfnamefont{M.~C.} \bibnamefont{Cross}}, \bibnamefont{and}
  \bibinfo{author}{\bibfnamefont{P.~F.} \bibnamefont{Fischer}},
  \bibinfo{journal}{Physical Review E}
  \textbf{\bibinfo{volume}{66}}(\bibinfo{number}{4}), \bibinfo{pages}{046210}
  (\bibinfo{year}{2002}).

\bibitem[{\citenamefont{Paul et~al.}(2001)\citenamefont{Paul, Cross, and
  Fischer}}]{Paul01}
\bibinfo{author}{\bibfnamefont{M.~R.} \bibnamefont{Paul}},
  \bibinfo{author}{\bibfnamefont{M.~C.} \bibnamefont{Cross}}, \bibnamefont{and}
  \bibinfo{author}{\bibfnamefont{P.~F.} \bibnamefont{Fischer}},
  \bibinfo{journal}{Physical Review Letters}
  \textbf{\bibinfo{volume}{87}}(\bibinfo{number}{15}), \bibinfo{pages}{154501}
  (\bibinfo{year}{2001}).

\bibitem[{\citenamefont{Paul et~al.}(2003)\citenamefont{Paul, Chiam, Cross,
  Fischer, and Greenside}}]{Paul03}
\bibinfo{author}{\bibfnamefont{M.~R.} \bibnamefont{Paul}},
  \bibinfo{author}{\bibfnamefont{K.-H.} \bibnamefont{Chiam}},
  \bibinfo{author}{\bibfnamefont{M.~C.} \bibnamefont{Cross}},
  \bibinfo{author}{\bibfnamefont{P.~F.} \bibnamefont{Fischer}},
  \bibnamefont{and} \bibinfo{author}{\bibfnamefont{H.~S.}
  \bibnamefont{Greenside}}, \bibinfo{journal}{Physica D}
  \textbf{\bibinfo{volume}{184}}, \bibinfo{pages}{114} (\bibinfo{year}{2003}).

\bibitem[{\citenamefont{Paul et~al.}(2004)\citenamefont{Paul, Chiam, Cross, and
  Fischer}}]{Paul04}
\bibinfo{author}{\bibfnamefont{M.~R.} \bibnamefont{Paul}},
  \bibinfo{author}{\bibfnamefont{K.-H.} \bibnamefont{Chiam}},
  \bibinfo{author}{\bibfnamefont{M.~C.} \bibnamefont{Cross}}, \bibnamefont{and}
  \bibinfo{author}{\bibfnamefont{P.~F.} \bibnamefont{Fischer}},
  \bibinfo{journal}{Physical Review Letters}
  \textbf{\bibinfo{volume}{93}}(\bibinfo{number}{6}), \bibinfo{pages}{064503}
  (\bibinfo{year}{2004}).

\bibitem[{\citenamefont{Fletcher}(1991)}]{Fletcher91}
\bibinfo{author}{\bibfnamefont{C.~A.~J.} \bibnamefont{Fletcher}},
  \emph{\bibinfo{title}{Computational Techniques for Fluid Dynamics}},
  vol.~\bibinfo{volume}{I} of \emph{\bibinfo{series}{Springer Series in
  Computational Physics}} (\bibinfo{publisher}{Springer-Verlag},
  \bibinfo{address}{Berlin}, \bibinfo{year}{1991}), \bibinfo{edition}{2nd} ed.

\bibitem[{\citenamefont{Tannehill et~al.}(1997)\citenamefont{Tannehill,
  Anderson, and Pletcher}}]{Tannehill97}
\bibinfo{author}{\bibfnamefont{J.~C.} \bibnamefont{Tannehill}},
  \bibinfo{author}{\bibfnamefont{D.~A.} \bibnamefont{Anderson}},
  \bibnamefont{and} \bibinfo{author}{\bibfnamefont{R.~H.}
  \bibnamefont{Pletcher}}, \emph{\bibinfo{title}{Computational Fluid Mechanics
  and Heat Transfer}} (\bibinfo{publisher}{Taylor and Francis},
  \bibinfo{address}{New York}, \bibinfo{year}{1997}), \bibinfo{edition}{2nd}
  ed.

\bibitem[{\citenamefont{Fischer and Mullen}(2001)}]{Fischer01}
\bibinfo{author}{\bibfnamefont{P.~F.} \bibnamefont{Fischer}} \bibnamefont{and}
  \bibinfo{author}{\bibfnamefont{J.~S.} \bibnamefont{Mullen}},
  \bibinfo{journal}{Comptes Rendus de {l'Acad\'emie} des Sciences Paris,
  {S\'erie} I, Analyse {Num\'erique}} \textbf{\bibinfo{volume}{332}},
  \bibinfo{pages}{265} (\bibinfo{year}{2001}).

\bibitem[{\citenamefont{Bohr et~al.}(1998)\citenamefont{Bohr, Jensen, Paladin,
  and Vulpiani}}]{Bohr98}
\bibinfo{author}{\bibfnamefont{T.}~\bibnamefont{Bohr}},
  \bibinfo{author}{\bibfnamefont{M.~H.} \bibnamefont{Jensen}},
  \bibinfo{author}{\bibfnamefont{G.}~\bibnamefont{Paladin}}, \bibnamefont{and}
  \bibinfo{author}{\bibfnamefont{A.}~\bibnamefont{Vulpiani}},
  \emph{\bibinfo{title}{Dynamical Systems Approach to Turbulence}}
  (\bibinfo{publisher}{Cambridge University Press},
  \bibinfo{address}{Cambridge}, \bibinfo{year}{1998}).

\bibitem[{\citenamefont{Shlesinger et~al.}(1995)\citenamefont{Shlesinger,
  Zaslavsky, and Frisch}}]{Shlesinger95}
\bibinfo{editor}{\bibfnamefont{M.~F.} \bibnamefont{Shlesinger}},
  \bibinfo{editor}{\bibfnamefont{G.~M.} \bibnamefont{Zaslavsky}},
  \bibnamefont{and} \bibinfo{editor}{\bibfnamefont{U.}~\bibnamefont{Frisch}},
  eds., \emph{\bibinfo{title}{L\'evy Flights and Related Topics in Physics}}
  (\bibinfo{publisher}{Springer}, \bibinfo{year}{1995}).

\bibitem[{\citenamefont{Egolf et~al.}(1998)\citenamefont{Egolf, Melnikov, and
  Bodenschatz}}]{Egolf98}
\bibinfo{author}{\bibfnamefont{D.~A.} \bibnamefont{Egolf}},
  \bibinfo{author}{\bibfnamefont{I.~V.} \bibnamefont{Melnikov}},
  \bibnamefont{and}
  \bibinfo{author}{\bibfnamefont{E.}~\bibnamefont{Bodenschatz}},
  \bibinfo{journal}{Physical Review Letters}
  \textbf{\bibinfo{volume}{80}}(\bibinfo{number}{15}), \bibinfo{pages}{3228}
  (\bibinfo{year}{1998}).

\end{thebibliography}

\end{document}